\documentclass[journal=ancac3,manuscript=article]{achemso}
\usepackage[utf8]{inputenc} 
\usepackage[T1]{fontenc}
\usepackage{upgreek}
\usepackage[version=3]{mhchem} 
\usepackage[font=small,labelfont=bf]{caption}
\usepackage{xcolor}

\def \MSE{Department of Materials Science and Engineering, National University of Singapore, 117575, Singapore}
\def \CADM{Centre for Advanced 2D Materials, National University of Singapore, 117546, Singapore} 
\def \IFIM{Institute for Functional Intelligent Materials, National University of Singapore, 117544, Singapore}
\def \CHEM{Department of Chemistry, National University of Singapore, 117543, Singapore}
\def \JAPAN{Research Center for Electronic and Optical Materials, National Institute for Materials Science, 1-1 Namiki, Tsukuba 305-0044, Japan}
\def \JAPANN{Research Center for Materials Nanoarchitectonics, National Institute for Materials Science, 1-1 Namiki, Tsukuba 305-0044, Japan}

\author{Z.~Qiu}
\affiliation{\MSE}
\alsoaffiliation{\IFIM}

\author{K.~Vaklinova}
\affiliation{\CADM}

\author{P.~Huang}
\affiliation{\MSE}
\alsoaffiliation{\IFIM}

\author{M.~Grzeszczyk}
\affiliation{\MSE}
\alsoaffiliation{\IFIM}

\author{K.~Watanabe}
\affiliation{\JAPAN}

\author{T.~Taniguchi}
\affiliation{\JAPANN}

\author{K.~S.~Novoselov}
\affiliation{\MSE}
\alsoaffiliation{\IFIM}

\author{J.~Lu}
\affiliation{\CADM}
\alsoaffiliation{\CHEM}
\email{chmluj@nus.edu.sg}

\author{M.~Koperski}
\affiliation{\MSE}
\alsoaffiliation{\IFIM}
\email{msemaci@nus.edu.sg}

\title{Atomic and electronic structure of defects in hBN: enhancing single-defect functionalities.}


\begin{document}

\begin{abstract}
\textbf{Defect centers in insulators play a critical role in creating important functionalities in materials: prototype qubits, single-photon sources, magnetic field probes, and pressure sensors. These functionalities are highly dependent on their mid-gap electronic structure and orbital/spin wave-function contributions. However, in most cases, these fundamental properties remain unknown or speculative due to the defects being deeply embedded beneath the surface of highly resistive host crystals, thus impeding access through surface probes. Here, we directly inspected the atomic and electronic structures of defects in thin carbon-doped hexagonal boron nitride (hBN:C) using scanning tunneling microscopy (STM) and scanning tunneling spectroscopy (STS). Such investigation adds direct information about the electronic mid-gap states to the well-established photoluminescence response (including single photon emission) of intentionally created carbon defects in the most commonly investigated van der Waals insulator. Our joint atomic-scale experimental and theoretical investigations reveal two main categories of defects: 1)~single-site defects manifesting as donor-like states with atomically resolved structures observable via STM, and 2)~multi-site defect complexes exhibiting a ladder of empty and occupied mid-gap states characterized by distinct spatial geometries. Combining direct probing of mid-gap states through tunneling spectroscopy with the inspection of the optical response of insulators hosting specific defect structures holds promise for creating and enhancing functionalities realized with individual defects in the quantum limit. These findings underscore not only the versatility of hBN:C as a platform for quantum defect engineering but also its potential to drive advancements in atomic-scale optoelectronics.} 

\end{abstract}
\section{Introduction}
Defect centers found or created in insulating crystals, like diamond, metal oxides, silicon carbide, or gallium nitride, can modify the fundamental electronic properties of the host material and lead to new phenomena. This has resulted in a variety of functionalities, such as testbed quantum systems in the form of qubits or entangled centers \cite{Pla2012}, photoluminescence markers \cite{PL_markers}, sensors of pressure or local magnetic fields in nanoscale \cite{defect_sensing}, lasing media \cite{defect_lasing}, or single photon sources \cite{single_photon_defect}. However, the development of these functionalities has primarily been restricted to bulk crystals with defect centers buried deep beneath the crystal surface. This has made it challenging to study the defect-related characteristics in the band structure of the host material via local surface probes, thereby preventing access to comprehensive information about the atomic and electronic structure of specific defects. These properties of defects were rather inferred speculatively through other methods, such as optical spectroscopy, which was limited by spatial resolution and sensitivity to transitions only between electronic states fulfilling electric dipole, spin, and momentum selection rules. However, in this study, we demonstrate that the advances in two-dimensional (2D) materials have made it possible to embed defects in atomically thin films and investigate them directly through scanning probes employing electrons tunneling through the insulating barrier. The combination of 2D insulators and scanning tunneling probes allows for the direct visualization of the mid-gap ladders of discrete energy states and their corresponding spatial features.

Directly applying scanning tunneling microscopy (STM) to investigate the bulk insulator surface is challenging, primarily stemming from its large band gap and high contact resistance. To overcome this limitation, alternative schemes were developed by incorporating graphene or graphite\cite{lyu2023gate,qiu2019giant,qiu2021visualizing}, including tunneling into a semi-metallic graphene monolayer deposited on bulk hBN crystal \cite{wong2015characterization}. In such cases, electrically active defects in the insulator become discernable in STM by imposing the local Coulomb potential on the electrons in the semi-metallic graphene. However, the microscopic information about the atomic structure and the energy landscape of the mid-gap defect states in hBN remained inaccessible since the tunneling current was predominantly contributed by the electronic states of semi-metallic graphene. Here, we investigated tunneling through a thin hexagonal boron nitride (hBN) film toward the bulk graphite substrate, inverting the previous heterostructure architectures. The thickness of the hBN film was limited to no more than three layers, ensuring a notable tunneling conductance indicated by prior research\cite{britnell2012electron}. The thin hBN film was isolated through mechanical exfoliation from a bulk crystal modified by annealing in a graphite furnace \cite{TANIGUCHI2007525}. This procedure was found to induce radiative centers with well-established spectral characteristics that were revealed by their photoluminescence response \cite{koperski2020midgap}. This enabled the systematic study of the optical properties of defects, leading to the development of single photon emitters \cite{dielectric_sens}, the sensitivity of the dielectric environment in the single-defect limit \cite{dielectric_sens}, and vertical tunneling light emitting diodes utilizing intra-defect optical transitions \cite{grzeszczyk2023electrical}. The characteristics of the optical resonances were compared with the theoretically predicted optical excitations in quantum-embedding schemes \cite{dielectric_sens}, leading to early proposals to attribute photoluminescence resonances to defects of specific atomic structures based on carbon substitutions and vacancies \cite{carbon_vacancy}. To achieve a fundamental understanding of the link between atomic structures, electronic band structures, and optical response of defects, we investigated the carbon-doped hBN (hBN:C) that has already been used to develop functionalities at a single-defect level. 

The present study sheds light on the defect properties in hBN and its potential as a platform for defect engineering. The development of tools for characterizing the atomic and electronic structures of individual defects\cite{fang2023atomically}, as demonstrated here through a combination of STM, STS, and optical spectroscopy, is pivotal for further advancements in various domains including single photon emission \cite{tran2016quantum, koperski2018single, koperski2021towards}, sensing capabilities \cite{dielectric_sens}, integration of quantum light with optoelectronic devices \cite{grzeszczyk2023electrical} and photonic structures.

\section{Results and Discussion}

\begin{figure}[h!]
\centering
    \includegraphics[width=\textwidth]{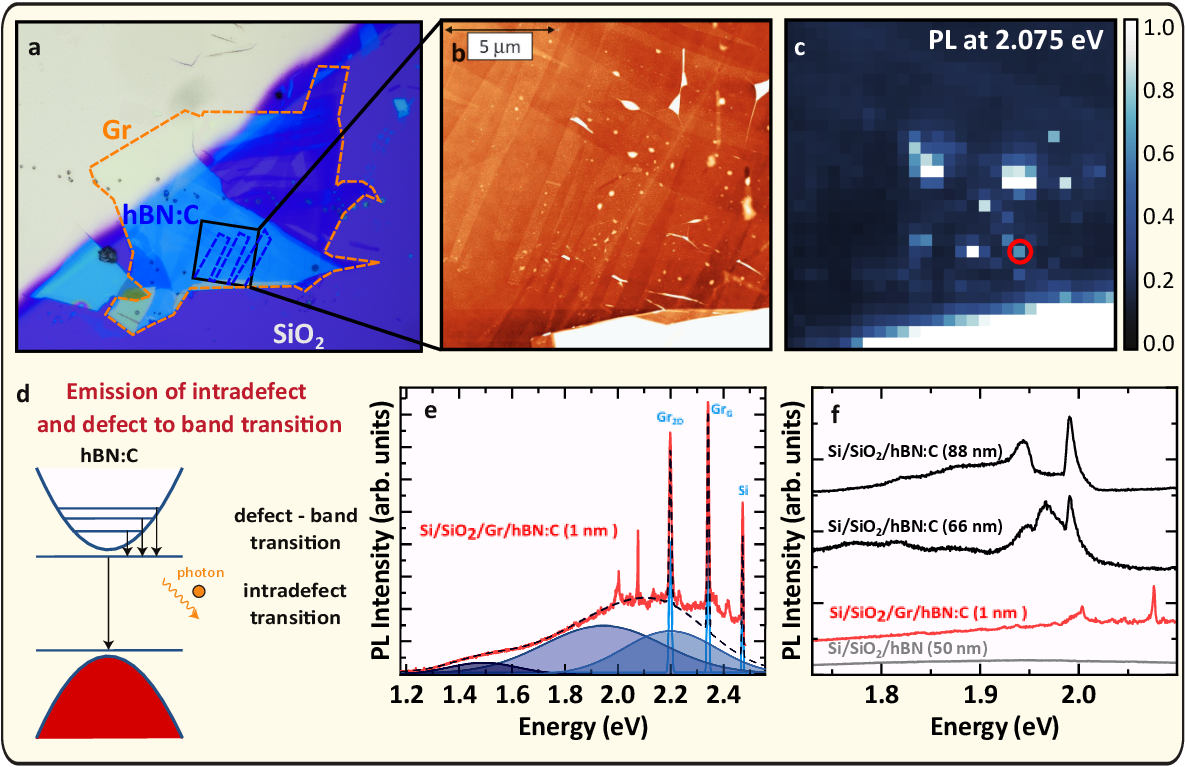}
    \caption{\textbf{Photoluminescence (PL) spectroscopy of Si/SiO$_2$/graphite/hBN:C device.} \textbf{(a)} The optical image of the sample presents the region with graphite/hBN:C heterostructure. \textbf{(b)} The atomic force microscopy image enabled the identification of three layers hBN:C film. \textbf{(c)} The map of integrated PL intensity at the energy 2.075~eV revealed localized emission from hBN:C. Maps \textbf{(b)} and \textbf{(c)} correspond to the same area of the sample. \textbf{(d)} A diagram demonstrating optical transitions involving mid-gap defect states and the electronic sub-bands of the host material. \textbf{(e)} A representative low-temperature optical spectrum from the location marked with a circle in the PL map in \textbf{(e)}. The optical response comprises Raman scattering resonances from graphite (2D and G bands), silicon (Si), and PL signal from hBN:C in the form of narrow resonances and broad bands. The broadband component was deconvoluted into three bands as highlighted by the Gaussian peaks. \textbf{(f)} The comparison between PL spectra measured for several hBN and hBN:C films characterized by varied thickness.} 
    \label{fig:1} 
\end{figure}

Enabling STM and STS characterization of atomically thin hBN required a sample of a specifically designed structure. To achieve this, we constructed Si/SiO$_2$/graphite/hBN:C heterostructures, where the (semi-)metallic graphite substrate was contacted by a gold electrode, facilitating electrical grounding and a convenient STM tip approach. The fabricated sample underwent extended annealing at around 310$^{\circ}$~C in a high vacuum environment, typically overnight, to achieve sufficient surface cleanliness for atomically-resolved STM imaging. The optical and atomic force microscopy (AFM) images of our sample with three layers of hBN:C are presented in \textbf{Fig.~\ref{fig:1}(a,b)}, while detailed information on the fabrication process can be found in the \textbf{Methods} section. The inspection of the optical response of this heterodevice was consistent with previous observations of radiative centers in hBN:C. Spatial maps of photoluminescence intensity, monitored at a specific energy of 2.075 eV corresponding to single photon emitters in thin layers of hBN:C\cite{dielectric_sens}, revealed localized emission originating only from regions covered by hBN:C film as seen in \textbf{Fig.~\ref{fig:1}(c)}. Generally, the introduction of mid-gap defect levels can activate different types of radiative processes. Transitions between the electronic bands of the host crystals and the defect states are typically characterized by broad emission bands indicative of a continuous dispersion of multiple conduction and/or valence sub-bands. The transitions between two defect states, in the limit of weak electron-phonon coupling, give rise to narrow photoluminescence resonances due to the discrete character of mid-gap levels. These two cases, i.e., defect to the band and intradefect radiative transitions, are schematically depicted in \textbf{Fig.~\ref{fig:1}(d)}. Their respective signatures (broadband emission and narrow resonances) coexist in the photoluminescence spectrum of Si/SiO$_2$/graphite/hBN:C sample shown in \textbf{Fig.~\ref{fig:1}(e)}. The narrow resonance at 2.075 eV exhibited an 85 meV blueshift relative to the resonances observable in bulk hBN:C films (see \textbf{Fig.~\ref{fig:1}(f)} for comparison between different hBN and hBN:C samples). Such modification of the emission energy was attributed previously to the effect of enhanced dielectric screening \cite{dielectric_sens}. Hence, the blueshift is expected in Si/SiO$_2$/graphite/hBN:C sample due to the proximity of thin hBN:C film to strongly screening metallic graphite substrate. 

\begin{figure}[ht!]
\centering
    \includegraphics[width=\textwidth]{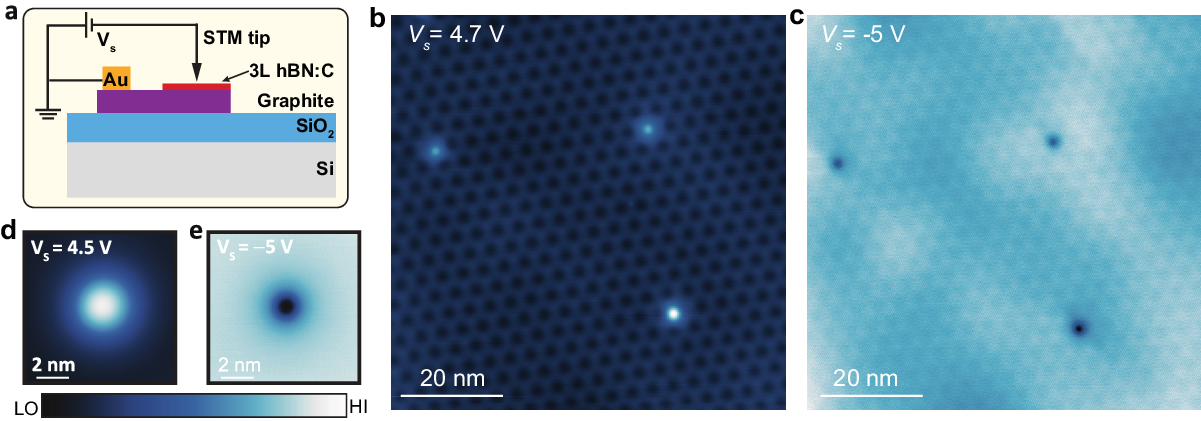}
    \caption{\textbf{Low-temperature ($\mathrm{\textbf{T=4.7}}$~K) scanning tunneling microscopy (STM) of the Si/SiO$_2$/graphite/hBN:C device.} \textbf{(a)} A diagram demonstrates the experimental configuration, where the graphite substrate was grounded via a golden electrode, and bias was applied to the STM tip. The large-scale STM images revealed the presence of multiple defect centers at \textbf{(b)} the positive sample bias $\mathrm{V_S}=4.7$~V, and \textbf{(c)} the negative sample bias $\mathrm{V_S}=-5.0$~V. The tunneling current was $\mathrm{I_t}=15$~pA, and  $\mathrm{I_t}=10$~pA for \textbf{(b)} and \textbf{(c)}, respectively. The individual defects emerged as bright protrusions at \textbf{(d)} positive sample bias $\mathrm{V_S}=4.5$~V  and as a dark halo at \textbf{(e)} negative sample bias $\mathrm{V_S}=-5.0$~V. The tunneling current was $\mathrm{I_t}=100$~pA, and  $\mathrm{I_t}=150$~pA for \textbf{(d)} and \textbf{(e)}, respectively.}
    \label{fig:fig2} 
\end{figure}

To directly investigate the defects in thin hBN:C film, we conducted the STM characterization of the sample in the experimental configuration presented schematically in \textbf{Fig.~\ref{fig:fig2}(a)}. For a large band gap system like hBN, we found it advantageous to initially perform STM imaging at a large bias V$_\text{S}$ of approximately 4.7 V. The large sample bias applied was comparable to the average work function of the tip and sample, positioning the STM operation within the bias range akin to the field emission regime. This strategy notably mitigated the risk of tip crashing while concurrently facilitating the mapping of the local electrostatic field of electrically active defects. \textbf{Fig.~\ref{fig:fig2}(b-c)} presented representative bias-dependent STM images with electrically active defects manifesting as protrusions/indentations for the positive/negative sample bias. A close examination revealed that the individual defect exhibited protrusions/indentations with a symmetric Gaussian-shaped extension characterized by a diameter of $1-2$~nm in \textbf{Fig.~\ref{fig:fig2}(d-e)}. Such an observation signified that these commonly observed defects are positively charged, creating a local electrostatic field that facilitates electron tunneling into empty states and suppresses hole tunneling into occupied states\cite{feenstra2002low,qiu2017resolving,telychko2022gate,fang2022electronic}.  Additional large-scale STM images were presented in S.~I.~Appendix in \textbf{Fig.~S1}, which indicated that positively charged defects were prevalent among electrically active defects. 

\begin{figure}[h!]
\centering
    \includegraphics[width=\textwidth]{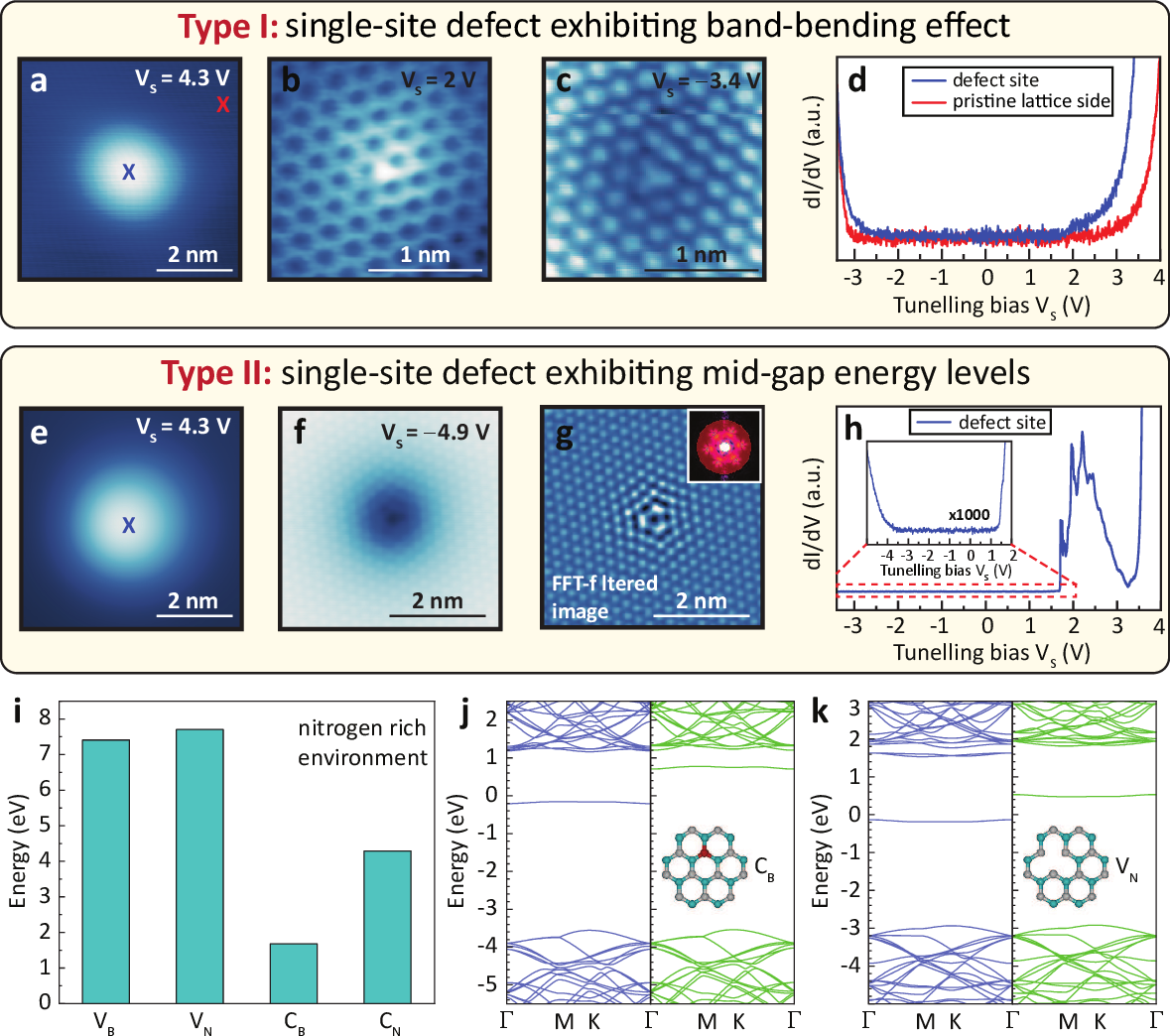}
    \caption{\textbf{Imaging and spectroscopy of single-site defects in hBN:C}. \textbf{(a-c)} Scanning tunneling microscopy (STM) images of the defect of the first type. The images were obtained at different sample biases: \textbf{(a)} $\mathrm{V_S}=4.3$~V, \textbf{(b)} $\mathrm{V_S}=2.0$~V, and \textbf{(c)} $\mathrm{V_S}=-3.4$~V. The tunneling current was $\mathrm{I_t}=30$~pA and the sample temperature was $\mathrm{T=77}$~K in all three cases. \textbf{(d)} Scanning tunneling spectroscopy (STS) transients are demonstrated for the pristine hBN lattice (red curve) and the defect of the first type (blue curve). The tip position is marked as red and blue crosses in panel \textbf{(a)}. \textbf{(e,f)} STM images of the defect of the second type. The images were obtained at different sample biases: \textbf{(e)} $\mathrm{V_S}=4.3$~V, and \textbf{(f)} $\mathrm{V_S}=-4.9$~V. The tunneling current was $\mathrm{I_t}=500$~pA and the sample temperature was $\mathrm{T=4.7}$~K in both cases. The single vacancy character of this defect was evident in \textbf{(g)} an image with a subtracted non-periodic contribution achieved via fast Fourier transform (FFT) filtering. The FFT image is demonstrated in the inset of \textbf{(g)}. \textbf{(h)} The STS transient for the defect of the second type was measured at the location marked by a blue cross in panel \textbf{(e)}. \textbf{(i)} The formation energy of boron vacancy ($\mathrm{V_B}$), nitrogen vacancy ($\mathrm{V_N}$), carbon substitution for boron ($\mathrm{C_B}$), and carbon substitution for nitrogen ($\mathrm{C_N}$) calculated in the framework of density functional theory (DFT). The DFT band structures of \textbf{(j)} $\mathrm{C_B}$ and \textbf{(k)} $\mathrm{V_N}$.}
    \label{fig:fig3} 
\end{figure}

Out of the population of positively charged defects, a subset of individual centers were selected at random for a detailed inspection of their properties. Atom-resolved STM images, summarised in \textbf{Fig.~S2}, served as the basis for this examination. Based on the symmetry observed in atom-resolved STM images, the examined defects could be categorized into two main groups: single-site defects characterized by a symmetric lattice, and multi-site defects characterized by an asymmetric lattice and a complex interplay between atomic and electronic structures. \textbf{Fig.~\ref{fig:fig3}(a-d)} illustrated a representative type of positively charged single-site defects, referred to as type I defects. The STM image acquired at V$_\text{S}$ = 4.3~V (\textbf{Fig.~\ref{fig:fig3}(a)}), revealed an extended Coulomb potential. At a lower bias of V$_\text{S}$ = 2~V, the defect center exhibited a significantly brightened dot feature (\textbf{Fig.~\ref{fig:fig3}(b)}). The surrounding lattice remained close to the crystallographic structure of pristine hBN, suggesting that the type I defect is related to a single heteroatom substitution. This interpretation is further supported by the STM image at V$_\text{S}$ = $- 3.4$~V, which unambiguously revealed a single-atom site defect center, as seen in \textbf{Fig.~\ref{fig:fig3}(c)}. Given the significant concentration of carbon in the hBN:C crystals identified previously via secondary ion mass spectroscopy \cite{carbon_doping}, the heteroatom substitution can be attributed to a carbon atom. Such impurity appeared systematically in this sample, and further examples of defects with analogous characteristics are presented in S.~I.~Appendix in \textbf{Fig.~S2}. Determining whether the substitution occured at a nitrogen or boron site could not be realized exclusively relying on STM imaging, which was analyzed based on data presented in \textbf{Fig.~S3}. Identifying the heteroatom substitutional position in the lattice was instead supported by \textit{dI/dV} spectroscopic measurement, as shown in \textbf{Fig.~\ref{fig:fig3}(d)}. The tunneling spectroscopy for pristine three-layer hBN region revealed a band gap of 6.20$~\pm~$0.29~eV via the inspection of the onsets indicative of the valence and the conduction band edges \cite{ugeda2014giant}. The detailed procedure for establishing these onsets is presented in \textbf{Fig.~S4}. In contrast, the tunneling \textit{dI/dV} spectrum corresponding to a defect site of the first type exhibited a lower onset of the conduction band minimum (CBM). The reduction of the CBM energy can be attributed to a significant downward band bending caused by the positively charged defect. That requires a donor-like defect level mixing with the CBM state, which can only occur in the considered scenario through carbon substitution for boron ($\mathrm{C_B}^+$).

We also identified a sparsely occurring type II defect, despite its identical signature in STM image acquired at V$_\text{S}$~=~4.3~V, presented in \textbf{Fig.~\ref{fig:fig3}(e)}. The atomic structure of the second defect type was starkly different from the previous case, as shown in the STM image at V$_\text{S}$~=~$-4.9$~V in \textbf{Fig.~\ref{fig:fig3}(f)}. The distinct structural characteristics of this defect were discernible when a non-periodic contribution from the extended Coulomb potential was subtracted based on fast Fourier transform filtering. As demonstrated in \textbf{Fig.~\ref{fig:fig3}(g)}, a single-atom vacancy became apparent, giving rise to a defect center of D$_\mathrm{3h}$ symmetry. The emergence of this defect was accompanied by local stretching of the surrounding hBN lattice, which constitutes a differentiating factor between a vacancy and an impurity. The \textit{dI/dV} spectrum of the vacancy, presented in \textbf{Fig.~\ref{fig:fig3}(g)}, revealed the presence of a ladder of narrow resonances below the CBM. Therefore, the tunneling spectroscopy indicated that the vacancy acts as a donor-like defect coupled with phonon excitations, which will be discussed in detail later. These characteristics are consistent with the vacancy at a nitrogen site ($\mathrm{V_N}^+$).

The fundamental properties of $\mathrm{C_B}$ and $\mathrm{V_N}$ defects can be comprehended in the framework of density functional theory (DFT). We calculated the formation energies of the carbon impurities and vacancies in the hBN lattice, as shown in \textbf{Fig.~\ref{fig:fig3}(i)}. These energies can be regarded as indicators of the probability of defect occurrence in hBN lattices exposed to high-temperature conditions in an atmosphere promoting the incorporation of specific impurities. The general trend is that carbon impurities are more likely to form compared with vacancies, as creating a vacancy requires more energy due to the necessity of breaking three covalent bonds. These DFT-predicted trends aligned with the frequency of observing vacancies and impurities in the atom-resolved STM images summarized in \textbf{Fig.~S2}. 

The DFT-calculated band structure of a unit cell containing $\mathrm{C_B}$ (\textbf{Fig.~\ref{fig:fig3}(j)}) and $\mathrm{V_N}$ (\textbf{Fig.~\ref{fig:fig3}(k)}) defects provides insights into their optical and tunneling spectroscopic characteristics. $\mathrm{C_B}$ defect gives rise to a single defect level below the CBM due to an excess electron provided by carbon at the boron lattice site. The optical response of such defect can be related to transitions between the defect level and the conduction band, which constitutes a plausible origin of the broad emission bands observed in the photoluminescence spectra of the hBN:C device. For the positive charge state of the defect, the excess electron can relax to the metallic state in the graphite substrate, thereby the ionized $\mathrm{C_B}^+$ exhibits the same number of electrons as boron in the neutral state. Such electronic configuration can favor the band-bending effect revealed by the STS data. The band structure characterizing the $\mathrm{V_N}$ defect is more intricate due to the presence of three broken bonds that give rise to three defect levels. One of the levels is located below the CBM, with the energy separation between the defect level and the band edge larger than in the case of $\mathrm{C_B}$ defect. The other two defect levels are located within the conduction band. In such an energy landscape, the mid-gap defect levels are expected to be better decoupled from the electronic bands of hBN, consistent with the observed emergence of narrow resonances below the CBM in the STS data. Although the optical response of the $\mathrm{V_N}$ defect could in principle combine the defect to the band and intradefect transitions, the predicted emission energy based on many-body models applied to intra-$\mathrm{V_N}$ transitions does not correspond to the single photon emitters observed so far in thin layers of hBN:C \cite{dielectric_sens}.

\begin{figure}[h!]
\centering
    \includegraphics[width=.88\textwidth]{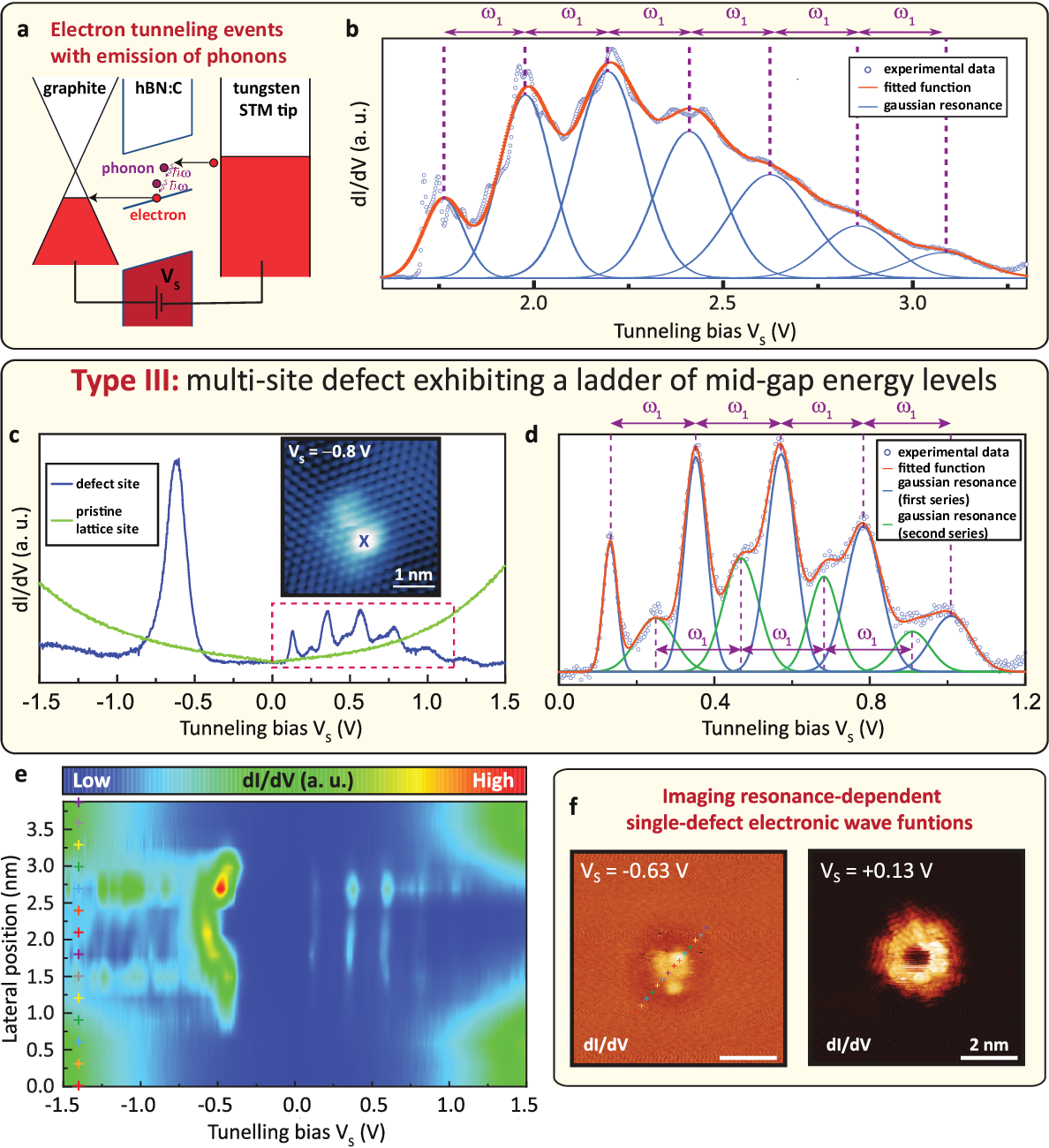}
    \caption{\textbf{Electron-phonon coupling and defects with a ladder of mid-gap states revealed by STS characterization.} \textbf{(a)} 
The diagram presents an inelastic tunneling process via a defect state assisted by multiple phonons. \textbf{(b)} A small-scale \textit{dI/dV} spectrum of the defect of the second type exhibited a series of resonances, replicated by fitting seven equidistant Gaussian peaks with a splitting of $\hbar\omega_1=220\pm7$~meV. \textbf{(c)} \textit{dI/dV} spectra for a pristine hBN lattice (green) and the defect of a third type (blue). The inset demonstrates the STM image of the defect of the third type under sample bias  $\mathrm{V_S}=-0.8$~V, and tunneling current $\mathrm{I_t}=30$~pA. The sample temperature was $\mathrm{T=77}$~K. \textbf{(d)} A small-scale \textit{dI/dV} spectrum for the defect of the third type, replicated by two subsets of equidistant Gaussian peaks with equal splitting of $\hbar\omega_1=220\pm7$~meV.  \textbf{(e)} Spatially resolved \textit{dI/dV} spectra across the defect of the third type. \textbf{(f)} The \textit{dI/dV} maps of the defect of the third type measured at different sample biases: $\mathrm{V_s}=-0.63$~V and $\mathrm{V_s}=0.13$~V. The corresponding position of the tip is marked by “+” symbols in panels \textbf{(e,f)}.} 
    \label{fig:fig4} 
\end{figure}

The tunneling spectra exhibited resonances that were not restricted to mid-gap defect states exclusively.  These resonances can also arise from tunneling events that are facilitated by the emission of phonons into the hBN lattice. This gives rise to replicated resonances at higher $\mathrm{V_S}$ values, as illustrated in \textbf{Fig.~\ref{fig:fig4}(a)}. The phonon excitations coupled to electrons occupying a defect state can often be modeled as harmonic oscillators. This model predicts equidistant resonances separated by the energy of a phonon. The mid-gap resonances observed for the $\mathrm{V_N}^+$ defect were found to be consistent with this scenario. Specifically, they were reproduced by seven Gaussian peaks split by $\hbar \omega_1 = 220 \pm 7$~meV, as demonstrated in \textbf{Fig.~\ref{fig:fig4}(b)}. However, shifts in the position of resonances induced by the electric field originating from the STM tip often result in an overestimation of the phonon energy in STS experiments, as demonstrated in \textbf{Fig.~S5}. Therefore, it is plausible that the dominant electron-phonon coupling mechanism is related to the Raman-active optical phonon mode in hBN at 195 meV, which corresponds to a bond-stretching vibrational mode \cite{Serano2007}. Furthermore, the lowest energy resonance at $\mathrm{V_S=1.75~V}$ exhibited a fine structure that could be reproduced by considering replicas of two phonon modes of the energy: $\hbar \omega_2 = 20 \pm 1$~meV and $\hbar \omega_3 = 30 \pm 3$~meV, as shown in \textbf{Fig.~S6}. Such low-energy vibrational modes coupled to an individual defect are likely to originate from local phonons that are spatially localized within the stretched lattice around the defect, as visualized by the STM images for defect $\mathrm{V_N}^+$. It is noteworthy that prior studies have underscored the important role of impurity states efficiently coupled to phonon modes in the optical response of single-photon emitters in hBN:C \cite{koperski2020midgap,wigger2019phonon}, which is evidenced by the common observation of phonon side bands accompanying the zero-phonon line, as can be seen in the PL spectra from hBN:C in \textbf{Fig.~\ref{fig:1}(f)}. This suggests that replicated resonances due to strong vibronic response in the tunneling spectrum may aid in the identification of potential defect candidates for the single-photon emission among the myriad of defects present within hBN:C. 

After elucidating the nature and electronic properties of positively charged single-site defects, we proceeded to explore multi-site defects characterized by multiple mid-gap states coupled with vibrational modes (\textbf{Fig.~\ref{fig:fig4}(c)}), which we denoted as type III defect. The observed characteristics suggest a more intricate atomic structure than a single-atom lattice modification. We hypothesize that the type III defect is associated with a complex of carbon impurities and/or vacancies, given the mechanism of reduced formation energies driven by the creation of molecular-like bonds in multi-site defects in hBN \cite{jara2021first,huang2012defect}. Above the Fermi level, the $dI/dV$ spectrum revealed the presence of two unoccupied defect states located at $\mathrm{V_S}=0.13$~V and $\mathrm{V_S}=0.24$~V strongly coupled to the phonon mode of the energy $\hbar \omega_1 = 220$~meV (\textbf{Fig.~\ref{fig:fig4}(d)}). Below the Fermi level, a prominent resonant peak appeared at $\mathrm{V_S}=-0.63$~V. The spatially-resolved STS data, presented in \textbf{Fig.~\ref{fig:fig4}(e)}, demonstrated that the resonance shifted to higher energy when the STM tip approached the defect center. Such an observation excluded the possibility of a tip-induced charging peak, as a charging peak would shift in the opposite direction\cite{qiu2017resolving}. Therefore, the physical origin of the resonance at $\mathrm{V_S}=-0.63$~V can be attributed to the occupied defect state. The occupied and empty defect states displayed wave functions of distinct geometry, as illustrated by bias-dependent \textit{dI/dV} mapping, presented in \textbf{Fig.~\ref{fig:fig4}(f)}. The occupied state at $\mathrm{V_S}=-0.63$~V exhibited an asymmetric wave function centered at the position of the defect, while the empty defect state at V$_\text{S}$=0.13~V revealed a wave function consisting of a dim center enclosed by a ring-like protrusion. While directly identifying the nature of the observed multi-site defects is challenging based on the limited information, the characteristics of the observed defect may suggest some possibilities. For instance, the presence of multiple in-gap states near the Fermi level with two unoccupied states and one occupied state coincided with the predicted electronic structure of $\mathrm{C_B V_N}$ defect \cite{dielectric_sens}, which was previously proposed as the origin of single photon emission in hBN films. This implication aligns with our direct observation of both constituent defects $\mathrm{C_B}$ and $\mathrm{V_N}$. However, at present, we cannot exclude more complicated defect centers, such as $\mathrm{C_B C_N V_N}$ or $\mathrm{C_B V_N V_B}$. 

\section{Conclusions}

In conclusion, we investigated the properties of individual defects in atomically thin hBN layers through defect-mediated tunneling of electrons toward a metallic substrate. A combination of STM, STS, and optical spectroscopy has proven to be a comprehensive toolset enabling the exploration of the energy landscape, wave functions, and optically active transitions arising due to fundamental defects in a van der Waals insulator. We found a defect population dominated by $\mathrm{C_B}^+$ centers. Given a relatively high mobility of single-site carbon substitution at evaluated temperature, intentionally introduced carbon impurities may combine further with other types of atomic defects to create thermodynamically stable complex defect structures, hosting a ladder of mid-gap states with sizeable electron-phonon coupling. The characterized electronic structure of the defect complex enables a many-body intra-defect transition that potentially constitutes the origin of single photon emission observed in thin layers of hBN. The direct access to mid-gap states in hBN presents the possibility of creating functionalities at a single defect level. We envisage that the next challenge for hBN-based research lies in developing scanning tunneling luminescence \cite{Lopez2023}, which would enable local electrical excitation of single-photon sources at the atomic limit.

\begin{acknowledgement}

This project was supported by the Ministry of Education (Singapore) through the Research Centre of Excellence program (grant EDUN C-33-18-279-V12, I-FIM), AcRF Tier 3 (MOE2018-T3-1-005) and MOE Tier 2 grants (MOE-T2EP10221-0005, MOE-T2EP10123-0004, and MOE-T2EP50122-0012), Agency for Science, Technology and Research (A*STAR) under MTC Individual Research Grants (M21K2c0113). This material is based upon work supported by the Air Force Office of Scientific Research and the Office of Naval Research Global under award number FA8655-21-1-7026. K.W. and T.T. acknowledge support from the JSPS KAKENHI (Grant Numbers 20H00354 and 23H02052) and World Premier International Research Center Initiative (WPI), MEXT, Japan.

\end{acknowledgement}

\section{Methods}

\textbf{Sample fabrication.} Graphite and hBN:C crystals were mechanically cleaved onto silicon wafers with 300~nm and 90~nm layers of \ce{SiO2}, respectively. Large-area graphite films were selected to act as conductive substrates for atomically thin hBN:C films. The hBN:C flakes were transferred onto the graphite substrate via a pickup technique. The hBN:C flakes of the desired thickness were identified by their optical contrast and atomic force microscopy. They were lifted from the Si/\ce{SiO2} wafer with a polydimethylsiloxane/polycarbonate stamp at 100$^\circ$C. Subsequently, the hBN:C flakes were released onto the graphite film together with the polycarbonate film at 180$^{\circ}$~C, which renders a high-quality interface between the two materials. The sample was then annealed at 180$^{\circ}$~C on a hot plate and washed in dichloromethane, acetone, and isopropanol to remove the polymer residue. Before the STM measurements, the sample was annealed in situ in an ultra-high vacuum at 310$^{\circ}$~C for 12~h to remove surface residues and adsorbates.

\noindent
\textbf{Optical characterization.} The photoluminescence (PL) spectra were measured in microscopic (1~$\upmu$m spot size) back-scattering geometry with 514~nm laser excitation. The sample was cooled down to 1.6~K through helium exchange gas. The sample was mounted on x-y-z piezo positioners that allowed for a PL mapping experiment. The PL signal from the sample was collected by a multimode fiber with 50~mm core diameter, dispersed by a 0.75~m spectrometer with 300~g/mm grating, and detected a by liquid-nitrogen-cooled charge-coupled device camera.

\noindent
\textbf{STM and STS measurements.} Our STM and STS measurements were conducted in the Createc LT-STM system with a base pressure lower than 10$^{-10}$~mbar at 4.7~K or 77~K. Performing STM/STS measurements at 4.7~K improved the energy resolution in tunneling spectroscopy at the cost of elevated risk of crashing the tip due to the reduced sample conductivity at low temperatures. However, the outcomes do not yield a qualitatively distinct result regarding the atomic and electronic structure of defects in hBN:C when compared to the STM/STS results obtained at 77~K. The tungsten tip was calibrated spectroscopically against the surface state of Au(111) substrate. All the \textit{dI/dV} spectra were measured through a standard lock-in technique with a modulated voltage of 3-10~mV at the frequency of 700-900~Hz.

\noindent
\textbf{DFT calculations.} The DFT calculations were performed using a generalized gradient approximation (GGA) to the exchange correlation functional proposed by Pedew, Burke and Ernzerhof (PBE)\cite{perdew1996generalized}. All calculations were spin-polarized and used the projector augmented wave (PAW) pseudo-potential supplied with the VASP code\cite{kresse1996efficiency,kresse1999ultrasoft}. A plane wave cut-off of 500~eV was used. Pristine single-layer hBN was first optimized for the lattice geometry using the conventional cell and a 21x21x1 Monkhurst-Pack reciprocal space grid with an energy tolerance of 0.01 eV. A vacuum spacing of 20~\AA~was used to separate periodic images of the single layer and the lattice vectors in a conjugate gradient approach. A 5x5x1 supercell was used for the calculations of the defect properties in hBN. The structures were fully optimized using the conjugate gradient algorithm until the residual atomic forces were smaller than 10~meV/\AA. A G-centreed 12×12 k-point sampling was used for the Brillouin-zone integration. The method for the calculation of the formation energy of different defects existing in various charge states at the N-rich condition was the same as in the Ref.\citenum{huang2012defect}

\bibliography{ref}

\end{document}